# A CONTINUOUS OPTIMIZATION APPROACH FOR FINANCIAL PORTFOLIO SELECTION UNDER DISCRETE ASSET CHOICE CONSTRAINTS

**Mahdi Moeini**

Braunschweig University of Technology, IBR, Algorithms Group,
Mühlenpfordtstr. 23, 38106 Braunschweig, Germany
moeini@ibr.cs.tu-bs.de

**Abstract:** In this paper we consider a generalization of the Markowitz's Mean-Variance model under linear transaction costs and cardinality constraints. The cardinality constraints are used to limit the number of assets in the optimal portfolio. The generalized model is formulated as a mixed integer quadratic programming (MIP) problem. The purpose of this paper is to investigate a continuous approach based on difference of convex functions (DC) programming for solving the MIP model. The preliminary comparative results of the proposed approach versus CPLEX are presented.

**Keywords:** portfolio selection problem, mixed integer programming, DC programming.

## 1 INTRODUCTION

Let us suppose that we are given a certain amount of money to invest. The investment must be done in a given set of assets or stocks. Each way of diversifying this amount of money between the given assets is called a *portfolio* [3]. The objective is to find a way to invest the money in the best possible way, which is called the optimal portfolio. This problem is known as the *portfolio selection problem* and it has been widely studied. Particularly, Markowitz [11] was one of the first researchers who provided a quantitative framework for finding the optimal portfolio. Markowitz [11] introduced the famous Mean-Variance (MV) model. The MV model is based on the expected return and the variance of returns between the assets [3]. The variance of returns is defined as the risk and, in this context; the objective of the portfolio selection problem consists of finding the set of portfolios offering the minimum level of risk for a given level of return. In order to find such portfolios, Markowitz proposes a convex quadratic programming (QP) model that is the MV model. This model has been widely used in practical applications. In spite of this fact, the standard MV model suffers from several inconveniences, for example, the MV model does not contain some practical constraints such as cardinality constraints, threshold constraints, or transaction costs functions. In fact, while an investor purchases or sells a stock, an extra charge will be made as the *transaction costs*. These costs must be taken into account in order to have realistic portfolio optimization models. There are different forms of the transaction costs functions: linear, piece-wise linear, step-wise linear functions, etc. The *cardinality constraints* limit the number of assets the optimal portfolio. The standard MV model is generalized by introducing these constraints [1-3]. The new model will be a mixed integer program (MIP) that is no more a convex programming problem. Due to the hardness of solving the MIP models, one needs to use local approaches that provide high quality solutions.

In this paper, we focus on solving the problem of portfolio selection under cardinality constraints in the presence of linear transaction costs that are proportional to the amount of the transactions. As the solution approach, a local deterministic method based on difference of convex functions (DC) programming and DC Algorithms (DCA) is used. This approach has been firstly introduced by Pham Dinh Tao in their preliminary form in 1985. They have been extensively developed since 1994 by Le Thi Hoai An and Pham Dinh Tao (see e.g. [7, 8, 12]). Due to successful application of the DC Algorithms for solving many large-scale mixed 0-1 programs (see, e.g., [4, 6, 8, 9]), a DC algorithm is developed for solving the generalized MV model. For testing the efficiency of proposed algorithm, we compare it with the results of the standard solver CPLEX.

The paper is organized as follows. After the introduction, we present in Section 2 the model of the portfolio selection problem under cardinality constraints and linear transaction costs functions. Section 3 deals with DC programming, the reformulation of the proposed model in term of a DC program, and a special realization of DC algorithms to the underlying portfolio selection problem. Section 4 is devoted to the experimental results and some conclusions are reported in Section 5.

## 2 PORTFOLIO SELECTION PROBLEM UNDER CARDINALITY CONSTRAINTS

First of all, let us remind the famous Markowitz's Mean-Variance model for the portfolio selection problem [3, 11]. Let $n$ be the number of available stocks, $r_i$ be the mean return of stock $i$ (for $i = 1,....,n$). $R \in \Re$ is the expected level of portfolio return and $Q$ is the variance-covariance matrix computed by using the historical returns of the assets. The decision variable $x_j$ is the proportional of the capital to be invested in the stock $j$. Using these notations, the standard Markowitz's Mean-Variance model is:

$$(P_{MV}): \quad \min\left\{x^t Q x : x^t r \geq R, \sum_{j=1}^{n} x_j = 1, x_j \geq 0\right\}.$$

This formulation is a simple convex quadratic program for which efficient algorithms are available. In this MV model, one minimizes the risk (i.e., $x^t Q x$) by ensuring the minimum level of portfolio return $R$.

In this paper, we study the generalized MV model by introducing realistic terms into the model. Particularly, we introduce the transaction costs and the cardinality constraints. The transaction costs are the amount of money that must be paid after each transaction (either purchasing or selling any stock). We suppose that the transaction costs are linear functions proportional to the amount of transactions. Furthermore, the cardinality constraints are introduced into the model to control the number of stocks representing the optimal portfolio. In order to define the cardinality constraints, we need to define the binary variables $z_j$ (for $j = 1,....,n$). We define $z_j = 1$ if and only if the stock $j$ is included in the optimal portfolio and $x_j \in [a_j, b_j]$, (where $0 \leq a_j \leq b_j \leq 1$ are lower and bounds, respectively), otherwise, $z_j$ will be equal to 0. Furthermore, we are going to use the following complementary notations:

- $c_b, c_s \in \Re^n$: the transactions costs vectors for purchasing and selling stocks, respectively. We suppose that the transaction costs are proportional to the amount of the transactions;
- $x_b, x_s \in \Re^n$: vectors of the purchasing and selling variables, respectively;
- $P \in \Re^n$: the current holding portfolio of the investor;
- $\bar{x} \in \Re^n$: the benchmark portfolio;
- $z \in \Re^n$: the vector of binary variables, that are used for formulating the cardinality constraints;
- $card$: the cardinality parameter defining the number of the stocks in the final portfolio.

The generalized model is as follows:

$(P_{card})$:

$$\min \quad (x - \bar{x})^t Q (x - \bar{x}) \tag{1}$$

Subject to:
$$(x - \bar{x})^t r - (c_b^t x_b + c_s^t x_s) \geq R, \quad (2)$$
$$P + x_b - x_s = x, \quad (3)$$
$$\sum_{j=1}^{n} x_j = 1, \quad (4)$$
$$\sum_{j=1}^{n} z_j = card, \quad (5)$$
$$a_j z_j \leq x_j \leq b_j z_j : j = 1,....,n, \quad (6)$$
$$z_j \in \{0,1\}: j = 1,....,n, \ x_b, x_s \geq 0. \quad (7)$$

By solving this problem, one minimizes the total risk associated with the portfolio to change the current position $P$ to the optimal portfolio $x^*$ by purchasing ($x_b$) some stocks or selling ($x_s$) them (constraint (3)). $\bar{x} \in \Re^n$ represents the benchmark portfolio that can be ignored by taking it equal to zero. It has no crucial role in our model. The current situation of the portfolio is defined by $P$, that can be taken equal to zero, as well. The total amount of paid transaction costs are computed by $(c_b^t x_b + c_s^t x_s)$. The model ensures that the optimal portfolio has an expected level of return denoted by $R$ after subtracting the transactions costs (constraint (2)). The constraint (4) means that the all amount of wealth must be invested in the stocks. The cardinality and bounding constraints are ensured by (5) and (6). The remaining constraints say which variables are continuous or binary.

It is well known that $(P_{card})$ is a Mixed Integer Program (MIP) that is an NP-hard problem. Due to this fact, one cannot use exact methods for solving this problem; particularly, if the dimension of the problem (i.e., $n$) is large. In the literature, different alternative methods have been proposed for solving the variants of MV model under cardinality constraints (see e.g., [2,3,5,9]). In this paper, we investigate a solution approach based on DC programming and DC algorithms for solving $(P_{card})$.

Before introducing the DC formulation of $(P_{card})$, a brief introduction to DC programming and DC algorithms is given in the following section.

## 3 SOLUTION METHOD VIA DC PROGRAMMING AND DC ALGORITHMS

### 3.1 DC Programming: A Short Introduction

In this section, we review some of the main definitions and properties of DC programming and DC Algorithms (DCA); where, *DC* stands for *difference of convex functions*.
Consider the following primal DC program
$$(P_{dc}): \quad \beta_p := \inf\{F(x) := g(x) - h(x) : x \in \Re^n\},$$
where $g$ and $h$ are convex and differentiable functions. $F(.)$ is a *DC function*, $g$ and $h$ are *DC components* of $F(.)$, and $g - h$ is called a *DC decomposition* of $F(.)$.
Let $C$ be a nonempty closed convex set and $\chi_C$ be the indicator function of $C$, i.e., $\chi_C(x) = 0$ if $x \in C$ and $+\infty$ otherwise. Then, one can transform the constrained problem
$$\inf\{g(x) - h(x) : x \in C\},$$
into the following unconstrained DC program
$$\inf\{f(x) := \varphi(x) - h(x) : x \in \Re^n\},$$

where $\varphi(x)$ is a convex function defined by $\varphi(x) := g(x) + \chi_C(x)$. Hence, without loss of generality, we suppose that the primal DC program is unconstrained and in the form of $(P_{dc})$. For any convex function $g$, its conjugate is defined by $g^*(y) := \sup\{\langle x, y \rangle - g(x) : x \in \Re^n\}$ and the dual program of $(P_{dc})$ is defined as follows

$$(D_{dc}): \quad \beta_d := \inf\{h^*(y) - g^*(y) : y \in \Re^n\},$$

One can prove that $\beta_p = \beta_d$ [12].

For a convex function $\theta$ and $x_0 \in \text{dom } \theta := \{x \in \Re^n : \theta(x_0) < +\infty\}$, the subdifferential of $\theta$ at $x_0$ is denoted by $\partial\theta(x_0)$ and is defined by

$$\partial\theta(x_0) := \{y \in \Re^n : \theta(x) \geq \theta(x_0) + \langle x - x_0, y \rangle, \forall x \in \Re^n\}$$

We note that $\partial\theta(x_0)$ is a closed convex set in $\Re^n$ and is a generalization of the concept of derivative.

For the primal DC program $(P_{dc})$ and $x^* \in \Re^n$, the *necessary* local optimality condition is described as follows

$$\partial h(x^*) \subset \partial g(x^*).$$

We are now ready to present the main scheme of the DC Algorithms (DCA) [12] that are used for solving the DC programming problems. The DC Algorithms (DCA) are based on local optimality conditions and duality in DC programming, and consist of constructing two sequences $\{x^l\}$ and $\{y^l\}$. The elements of these sequences are trial solutions for the primal and dual programs, respectively. In fact, $\{x^{l+1}\}$ and $\{y^{l+1}\}$ are solutions of the following convex primal program $(P_l)$ and dual program $(D_{l+1})$, respectively:

$$(P_l): \quad \inf\{g(x) - h(x^l) - \langle x - x^l, y^l \rangle : x \in \Re^n\},$$

$$(D_{l+1}): \quad \inf\{h^*(y) - g^*(y^l) - \langle y - y^l, x^{l+1} \rangle : y \in \Re^n\}$$

One must note that, $(P_l)$ and $(D_{l+1})$ are convexifications of $(P_{dc})$ and $(D_{dc})$, respectively, in which $h$ and $g^*$ are replaced by their corresponding affine minorizations. By using this approach, the solution sets of $(P_{dc})$ and $(D_{dc})$ are $\partial g^*(y^l)$ and $\partial h(x^{l+1})$, respectively. To sum up, in an iterative scheme, DCA takes the following simple form

$$y^l \in \partial h(x^l); \qquad x^{l+1} \in \partial g^*(y^l).$$

One can prove that the sequences $\{g(x^l) - h(x^l)\}$ and $\{h^*(y^l) - g^*(y^l)\}$ are decreasing, and $\{x^l\}$ (respectively, $\{y^l\}$) converges to a primal feasible solution (respectively, a dual feasible solution) satisfying the local optimality conditions. More details, on convergence properties and theoretical basis of the DCA, can be found in [12].

### 3.2   Reformulation of the problem

The model $(P_{card})$ is not in the form of a DC program. In order to reformulate $(P_{card})$, we use an exact penalty result presented in [10]. The process consists of formulating $(P_{card})$ in the form of a convex-concave minimization problem with linear constraints which is consequently a DC program. In order to simplify the notations, let us define

$$A := \left\{ \begin{array}{l} (x, x_b, x_s, z) \in \Re_+^{3n} \times [0,1]^n : \sum_{j=1}^n x_j = 1, (x-\bar{x})^t r - (c_b^t x_b + c_s^t x_s) \geq R, P + x_b - x_s = x, \\ \sum_{j=1}^n z_j = card, a_j z_j \leq x_j \leq b_j z_j : j = 1,\ldots, n \end{array} \right\}.$$

Using this notation, the $(P_{card})$ is transformed to

$$\min\{(x-\bar{x})^t Q(x-\bar{x}) : (x, x_b, x_s, z) \in A, z_j \in \{0,1\}: \forall j\} \quad (8)$$

Define the penalty function $\alpha(.)$ by $\alpha(x, x_b, x_s, z) := \sum_{j=1}^n z_j(1-z_j)$. Clearly, $\alpha(.)$ is a concave function with nonnegative values on $A$ and the feasible solutions' set of (8) can be written as

$$\{(x, x_b, x_s, z) \in A, z_j \in \{0,1\}: \forall j\} = \{(x, x_b, x_s, z) \in A, \alpha(x, x_b, x_s, z) \leq 0\}.$$

Consequently, (8) can be written as

$$\min\{(x-\bar{x})^t Q(x-\bar{x}) : (x, x_b, x_s, z) \in A, \alpha(x, x_b, x_s, z) \leq 0\} \quad (9)$$

Since $(x-\bar{x})^t Q(x-\bar{x})$ is convex and $A$ is a bounded polyhedral convex set, according to [10], there is $\theta_0 \geq 0$ such that for any $\theta > \theta_0$, the program (9) is equivalent to

$$(P_{card} - DC): \quad \min\{F := (x-\bar{x})^t Q(x-\bar{x}) + \theta\alpha(x, x_b, x_s, z) : (x, x_b, x_s, z) \in A\} \quad (10)$$

The function $F$ is convex in variables $x, x_b, x_s$ and concave in variables $z$. Hence, the objective function of $(P_{card} - DC)$ is a DC function. A natural DC formulation of the problem $(P_{card} - DC)$ is

$$g(x, x_b, x_s, z) := (x-\bar{x})^t Q(x-\bar{x}) + \chi_A(x, x_b, x_s, z) \text{ and } h(x, x_b, x_s, z) := \theta \sum_{j=1}^n z_j(z_j - 1),$$

where $\chi_A$ is the indicator function over $A$, i.e., $\chi_A(x, x_b, x_s, z) = 0$ if $(x, x_b, x_s, z) \in A$, and $+\infty$, otherwise.

### 3.3  A DC algorithm for solving $(P_{card} - DC)$

According to the general framework of DC algorithms, we first need computing a point in the subdifferential of the function $h$ defined by $h(x, x_b, x_s, z) := \theta \sum_{j=1}^n z_j(z_j - 1)$. This is done by:

$$(u^k, u_b^k, u_s^k, v^k) \in \partial h(x^k, x_b^k, x_s^k, z^k) \Leftrightarrow u^k = u_b^k = u_s^k = 0, \quad v^k = \theta(2z^k - 1). \quad (11)$$

Secondly, in order to compute $(x^{k+1}, x_b^{k+1}, x_s^{k+1}, z^{k+1}) \in \partial g^*(u^k, u_b^k, u_s^k, v^k)$, we need to solve the following *convex* quadratic program:

$$\min\{(x-\bar{x})^t Q(x-\bar{x}) - \langle (u^k, u_b^k, u_s^k, v^k), (x, x_b, x_s, z) \rangle : (x, x_b, x_s, z) \in A\} \quad (12)$$

To sum up, the DC algorithm for solving $(P_{card} - DC)$ can be described as follows:

**DC Algorithm for solving** $(P_{card} - DC)$

1) **Initialization**: Let $\varepsilon$ be a sufficiently small positive number, let $(x^0, x_b^0, x_s^0, z^0) \in \Re_+^{3n} \times [0,1]^n$, and set $k = 0$;
2) **Iterations**: For $k = 0,1,2,\ldots$, set $u^k = u_b^k = u_s^k = 0$, $v^k = \theta(2z^k - 1)$, and solve (12).
3) **Stopping criterion**: If $\|(x^{k+1}, x_b^{k+1}, x_s^{k+1}, z^{k+1}) - (x^k, x_b^k, x_s^k, z^k)\| \leq \varepsilon$, then stop, $(x^{k+1}, x_b^{k+1}, x_s^{k+1}, z^{k+1})$ is a solution, otherwise set $k \leftarrow k+1$ and go to the Step 2.

# 4 COMPUTATIONAL EXPERIMENTS AND RESULTS

The algorithm has been tested on two benchmark data sets that have been already used in [2, 3, 5]. These data sets correspond to weekly prices coming from the indices: *Hang Seng* in Hong Kong and *Dax 100* in Germany. The number $n$ of different assets is 31 and 85, respectively. We suppose that $a_j = 0.05$ and $b_j = 1.0$ for both indices. Furthermore, $\theta$ is set to be 2.0, $\varepsilon$ is equal to $10^{-6}$, $P_j = 0$ and $\bar{x}_j = 1/n$ (for $j = 1,...,n$), $c_b, c_s = 0.1\%$ of transaction (buying/selling), and finally the value of $R$ is chosen in a way to get feasible models. We have tested DCA and the standard IP solver IBM CPLEX for different values of the cardinality parameter *card*. A time limit of *1200* seconds has been set on the IP solver IBM CPLEX. In order to find a good initial solution for DCA, we first solve the relaxed problem of $(P_{card})$. The solution may not be integer, hence we round up each nonzero value to get an integer point.

In Tables 1 and 2, we give the results for two considered data sets. In these tables, the number of iterations of DCA, the computing time in seconds (CPU), and the solution values (Optimal Val.) obtained by each of the methods are presented.

*Table 1: The results for the index Hang Seng in Hong Kong.*

| card | CPLEX | | DC Algorithm (DCA) | | |
|---|---|---|---|---|---|
| | Optimal Val. | CPU(s.) | Optimal Val. | CPU(s.) | Iterations |
| 5  | 0.000080 | 4.031   | 0.000110 | 0.094 | 3 |
| 6  | 0.000062 | 10.297  | 0.000095 | 0.094 | 4 |
| 7  | 0.000052 | 29.500  | 0.000084 | 0.110 | 4 |
| 8  | 0.000043 | 54.485  | 0.000084 | 0.110 | 4 |
| 9  | 0.000038 | 107.860 | 0.000051 | 0.093 | 4 |
| 10 | 0.000033 | 154.546 | 0.000044 | 0.109 | 4 |
| 11 | 0.000029 | 140.562 | 0.000042 | 0.125 | 4 |
| 12 | 0.000026 | 48.235  | 0.000027 | 0.094 | 4 |
| 13 | 0.000022 | 21.141  | 0.000025 | 0.110 | 4 |
| 14 | 0.000020 | 9.906   | 0.000024 | 0.109 | 4 |
| 15 | 0.000018 | 3.094   | 0.000023 | 0.094 | 4 |

*Table 2*: *The results for the index DAX 100 in Germany.*

| card | CPLEX | | DC Algorithm (DCA) | | |
|---|---|---|---|---|---|
| | Optimal Val. | CPU(s.) | Optimal Val. | CPU(s.) | Iterations |
| 5  | 0.000071 | 1201.969 | 0.000114 | 0.343 | 4 |
| 6  | 0.000057 | 1201.157 | 0.000078 | 0.344 | 4 |
| 7  | 0.000050 | 1201.422 | 0.000072 | 0.360 | 4 |
| 8  | 0.000041 | 1201.297 | 0.000060 | 0.375 | 4 |
| 9  | 0.000037 | 1202.016 | 0.000056 | 0.344 | 4 |
| 10 | 0.000030 | 1201.500 | 0.000101 | 0.359 | 4 |
| 11 | 0.000029 | 1201.281 | 0.000068 | 0.360 | 4 |
| 12 | 0.000027 | 1201.282 | 0.000083 | 0.344 | 4 |
| 13 | 0.000026 | 1201.343 | 0.000050 | 0.359 | 4 |
| 14 | 0.000021 | 1201.110 | 0.000041 | 0.375 | 4 |
| 15 | 0.000020 | 1200.938 | 0.000038 | 0.359 | 4 |

The computational results show that DCA gives a good approximation of the optimal solution within a very short time. The running time is less than 1 second and the number of iterations is at most 4. It is interesting that the most of the values provided by DCA are exact until 4 or 5 digits after the point. When we compare the computational time that Cplex needs to find the solutions and the CPU time of the DCA, the achievements of the algorithm become more interesting.

## 5  CONCLUSIONS

In this paper, a new approach for solving the portfolio selection problem has been presented. Instead of the standard Markowitz Mean-Variance (MV) model, we have used an extension including the cardinality and bounding constraints. Furthermore, the extended model takes into account the linear transaction costs functions. The extended portfolio selection model is nonconvex and, consequently, very difficult to solve by existing algorithms. We have transformed the model to a DC program and developed a deterministic approach based on DC programming and DC algorithms (DCA). Preliminary numerical simulations show the efficiency of the proposed approach and its inexpensiveness in comparison to the standard IP solver of CPLEX. The good results make it possible to extend the work to larger dimensions and combining the DC algorithm with exact approaches in order to have a guarantee on the quality of the solutions. The work in these directions is currently in progress.